\documentclass[12pt]{article}

\usepackage[final]{graphicx}
\usepackage{amssymb}
\begin{document}

\author{M. Castelnovo, P. Sens,  and J.-F. Joanny\\  
\textit{Institut Charles Sadron (CNRS UPR $022$), 6 rue Boussingault,}\\
\textit{\ 67083
Strasbourg Cedex, France.}}
\title{Charge Distribution on Annealed Polyelectrolytes.}

\maketitle

\begin{abstract}
We investigate the equilibrium charge distribution along a single annealed polyelectrolyte chain under different conditions. The coupling between the conformation of the chain and the local charge distribution is described for various solvent qualities and salt concentration. In salt free solution, we find a slight charge depletion in the central part of the chain: the charges accumulate at the ends. The effect is less important if salt is added to the solution since the charge inhomogeneity is localized close to the chain ends over a distance of order of the Debye length. In the case of poor solvent conditions we find a different charging of beads and strings in the framework of the necklace model. This inhomogeneity leads to a charge instability and a first order transition between spherical globules and elongated chains.
\end{abstract}
%
%
%
\section{Introduction}
The study of charged polymers is nowadays a field of highly active
research \cite{barrat_joanny_gen} because both of their relations with biological systems and
of their practical applications. The main advantage compared to neutral polymers comes from
the ability of various polyelectrolytes to dissolve in polar solvents and in particular in
water. The polymer backbones are ionized and dissociate in the solution into polyions and free
counterions. It has been recognized that the distribution of the charges along the polyion
chain is an important parameter which allows to distinguish two types of polyelectrolytes
\cite{muller,dubin} : the charge distribution can be either quenched or annealed. For quenched
polyelectrolytes, the charges along the chain are in a frozen configuration as in polystyrene
sulfonate. The charge distribution depends on the initial chemistry  of the macromolecule but
remains fixed in the whole range of pH of the solution. On the other hand, the charge
distribution of polyacids or polybases strongly depends on pH and therefore the ionization
sites are able to move along the chain. By tuning the pH of the solution, the total charge of
the polymer can be monitored. One speaks of annealed or weak polyelectrolytes. This extra
degree of freedom for the charges leads to some new and non trivial behaviors. Zhulina et al.
\cite{zhulina_borisov} have shown for grafted layers of weak polyelectrolytes the existence of a
regime where the brush thickness is decreasing when the chain grafting density is increased.
This is mainly due to the fact that the net charge of a chain is not fixed but depends on its
local environment. Another interesting effect is the behavior of annealed polyelectrolytes in
poor solvent. Raphael and Joanny have predicted a first order phase transition from a compact
conformation of the chains to an extended state \cite{raphael_joanny}. This phenomenon
explains the sharp increase in the viscosity and the anomalous titration curves observed in
experiments \cite{muller,dubin}.

The purpose of the present paper is to investigate under various conditions the equilibrium
charge distribution along one isolated annealed polyelectrolyte chain, i.e. in the limit of
infinite dilution. This problem may seem very academic, but it is a first step towards a
more precise understanding of collective properties such as those predicted by Zhulina et al.:
in an annealed polyelectrolyte brush or in a related system such as a polyelectrolyte star,
one expects the charge distribution to be non-uniform along the chain. This work could also be
used to study related systems such as the self-assembly of weakly charged linear micelles,
where end effects are known to have a major influence on the size distribution \cite{vanderschoot}.

The first estimation of the charge distribution along a polyelectrolyte chain is due to
Berghold et al. \cite{berghold_seidel} using numerical simulations. We find qualitatively the
same important result: the charges accumulate at the chain ends where there is less
neighboring charges than in the middle of the chain. The charge distribution is
thus an end effect. However our model does not describe the computer simulations of 
ref.\cite{berghold_seidel} since, in their case, the chain conformation is almost unperturbed by the addition of electric charges. It means that the
chains used in the simulation are not in an asymptotic regime where one can use the concept of
electrostatic blobs. The charge distribution observed in \cite{berghold_seidel} would be
better described by a model where the neutral polymer conformation is imposed and where the
distribution of a few charges on the chain is then investigated. We will not proceed further
with such a model in this paper.

This paper is organized as follows. In section \ref{salt_free} we evaluate the charge
distribution in a salt-free solution, in an ideal (theta) solvent. We consider first the
effect of the finite length of the chain on the conformation of a polyelectrolyte with a
frozen charge distribution and then the charge distribution along an annealed polyelectrolyte.
We also generalize our results to the case of good solvents. In section \ref{salt}, 
we study the effect of added salt. Finally, we turn to the special case of a poor solvent, and
discuss the distribution of charges in the so-called necklace conformation. All the results
are discussed in the concluding section.

 \section{Charge distribution in a salt-free solution} \label{salt_free} \subsection{Conformation of a quenched
polyelectrolyte in theta solvent} 
\setcounter{equation}{0}\renewcommand{\theequation}{\thesubsection.\arabic{equation}}
\label{conformation} The charge distribution along a single
chain is an end effect: the charges are located at the minima of the electrostatic potential.
For an infinite chain, the average potential is constant along the chain; on the contrary, for
a finite chain, the charges at the ends have less neighbors than in the center, this leads to
inhomogeneities in the electrostatic potential. Our first step is to calculate the end effect
on the conformation of a chain with a frozen charge distribution. We then study the
coupling between the conformation and the charge distribution. For a quenched chain with
frozen charges, we expect a trumpetlike shape instead of a linear chain of constant
electrostatic blobs since the chain is more stretched at the center than at the ends (Fig.1).

We consider a flexible polyelectrolyte of $N$ monomers which carries $fN$ charges distributed
uniformly along the chain. The monomer length is denoted by $b$. The polymer backbone is
described in a continuous way and the position of a monomer in space is represented by
$\mathbf{r}(s)$; the first derivative $\frac {\partial \mathbf{r}}{\partial s}$ gives the
tension along the chain. We neglect the effect of the counterions in this part since we are
dealing with a single chain. When the charge distribution along the chain is annealed, the
chemical potential of the charges related to the pH of the solution is an important variable.
We present here an argument similar to the one given by Brochard for the shape of a tethered
polymer in a strong solvent flow \cite{brochard}. We write a force balance on each point of
the chain which amounts to neglecting the fluctuations around the so-called classical path of the
polymer (the most probable conformation). The force balance reduces to:
\begin{equation}
\label{f_balance1}
\frac{3kT}{b^{2}} \frac{d^{2}\mathbf{r}(s)}{ds^{2}}
=-ef\mathbf{E}(s)
\end{equation}
where $\mathbf{E}(s)$
is the average electric field created by the charges of the chain.
Eq.(\ref{f_balance1}) is valid at every point of the chain
except for the end points
where the force balance on a section of length $ds$ reads:
\begin{equation}
\label{f_balance1'}
\frac{3kT}{2b^{2}} \frac{d\mathbf{r}(s)}{d s}=-ef\mathbf{E}(s)ds
\end{equation}
because the chain is only stretched by one side at the end point.
Since the electrostatic force is proportional to $ds$,
the tension vanishes on the end point of the chain.
We now calculate the electric field
created by the polymer path $\mathbf{r}(s)$
in order to get a self-consistent equation for $\mathbf{r}(s)$. As we are using a continuous theory,
we have to introduce short distance cutoffs
to prevent the electric field from diverging.
The relevant length at position $s$ on the chain
is the local stretching blob size
\cite{brochard}:
\begin{equation}
\label{blob_def1}
\frac{dz}{ds}\equiv\frac{\xi (z(s))}{g(z(s))}=\frac{b^{2}}{\xi (z(s))}
\end{equation}
$z(s)$ being the polymer path
projected along the direction of stretching
and $g(z(s))$ the number of monomer
inside the blob centered at $z(s)$.
This way of introducing the blob size
is totally consistent with the definition of
the electrostatic blob, namely the length below which the electrostatic interactions are not relevant.
The closest charges to a test charge on the polymer
do not contribute significantly to the electric field at this point,
because of the natural highly fluctuating state
of the neutral polymer backbone.
This is shown more precisely in the appendix
with a model of chain under tension.
The linear charge density is $\frac{Q(z)}{\xi (z)}$
where $Q(z)=f(\frac{\xi(z)}{b})^{2}$
is the charge of the blob located at $z$.

The electric field is written therefore
as a sum over the contributions of all the blobs:
\begin{equation}
\label{elec_field1}
E(z)=\frac{fl_{B}kT}{eb^{2}}
\left\{
\int_{-L/2}^{z-\xi_{z}}dz_{1}
\frac{\xi_{z_{1}}}{(z-z_{1})^{2}}
-
\int_{z+\xi_{z}}^{L/2}dz_{1}
\frac{\xi_{z_{1}}}{(z-z_{1})^{2}}
\right\}
\end{equation}
$l_{B}=\frac{e^{2}}{4\pi\epsilon kT}$ is the Bjerrum length characterizing
the strength
of electrostatic interactions.
Eq.(\ref{f_balance1})
can be integrated over the monomer position $s$ to
give an integral equation for the blob size:
\begin{equation}
\label{inteq_blob1}
\frac{3kT}{2\xi_{z}}=
\frac{f^{2}l_{B}kT}{b^{4}}
\int_{z}^{L/2}dz_{2}\,\xi_{z_{2}}
\left\{
\int_{-L/2}^{z_{2}-\xi_{z_{2}}}dz_{1}
\frac{\xi_{z_{1}}}{(z_{2}-z_{1})^{2}}
-
\int_{z_{2}+\xi_{z_{2}}}^{L/2}dz_{1}
\frac{\xi_{z_{1}}}{(z_{2}-z_{1})^{2}}
\right\}
\end{equation}
In order to obtain a simple expression
for the blob size, we assume that $\xi_{z}$ is slowly varying with $z$ 
(assumption checked at posteriori), so that the blob size in the integrals
is chosen as its value at position $z$. This allows for an estimate of
the integrals of the right hand side of Eq.(\ref{inteq_blob1}):
\begin{eqnarray}
\label{inteq_blob2}
\frac{3kT}{2\xi_{z}^{3}}
& = &
\frac{uf^{2}}{b^{3}}
\Bigg\{
\int_{z}^{L/2-\xi_{M}}dz_{2}\,
\left[
\int_{-L/2}^{z_{2}-\xi_{z_{2}}}\frac{dz_{1}}{(z_{2}-z_{1})^{2}}
\,-\,
\int_{z_{2}+\xi_{z_{2}}}^{L/2}\frac{dz_{1}}{(z_{2}-z_{1})^{2}}
\right]
\nonumber\\
& & +
\int_{L/2-\xi_{M}}^{L/2}dz_{2}\,
\left[
\int_{-L/2}^{z_{2}-\xi_{M}}\frac{dz_{2}}{(z_{2}-z_{1})^{2}}
\right]
\Bigg\}
\end{eqnarray}
where $u\equiv l_B/b$, and the maximal blob size $\xi_{M}\equiv\xi(z=L/2-\xi_{M})$.

We finally obtain the blob size:
\begin{equation}
\xi_{z}=\frac{\xi_{\mathrm{scaling}}}
{\left[\ln{\left(\frac{(L/2)^{2}-z^{2}}{L\,\xi_{M}}\right)}+1\right]^
{1/3}}
\end{equation}
where $\xi_{\mathrm{scaling}}=\frac{b}{(uf^{2})^{1/3}}$ is
electrostatic blob size one gets from the scaling theory.
This expression holds for values of $z$
ranging from $-L/2+\xi_{M}$
to $L/2-\xi_{M}$.
The parameters $L$ and $\xi_{M}$
are determined self-consistently by the following equations:

\begin{eqnarray}
\label{def_length}
Nb^{2} & = & 2\,\int_{0}^{L/2-\xi_{M}}dz\,\xi_{z}
\\
\label{def_ximax}
\xi_{M} & = &
\frac{\xi_{\mathrm{scaling}}}{\left[\ln{\left(\frac{L-\xi_{M}}{L}\right)}+1\right]^{1/3}}
\end{eqnarray}
Eq.(\ref{def_length}) is the definition
of the local stretching blob size
and Eq.(\ref{def_ximax}) the definition of the maximal blob size.

This functional form of the blob size
is consistent with the assumption of slow variations with $z$:
the end effect on the conformation of the polyelectrolyte is weak.
We also note that
we recover the variational result of de Gennes et al.
\cite{degennes_pincus}
for the logarithmic correction of the total length of the chain
if we neglect the $z$ dependence of the logarithm:
\begin{equation}
L
\sim
\frac{N}{g}\,\xi_{(z\sim\mathrm{cst})}
\sim
\frac{Nb^{2}}{\xi_{\mathrm{scaling}}}
\,
\left[
\ln{\left(\frac{L}{\xi_{\mathrm{scaling}}}\right)}
\right]^{1/3}
\end{equation}
Although the approximation made to obtain the blob size
looks quite rough, we believe that we catch in this way the main features of the trumpetlike shape of 
the chain. One can show that the same result could also be derived from the free energy Eq.(\ref{free_energy}), given in the appendix.
%
\subsection{Charge distribution on an annealed polyelectrolyte}
\label{ch_distribution}
\setcounter{equation}{0}
We now turn to the annealed charge distribution
\cite{raphael_joanny}.
The total charge of the chain is not fixed but it
is imposed by the pH of the solution
because of dissociation and recombination of ion pairs along the chain.
The ionization sites can therefore move freely.
For quenched polyelectrolytes, in the previous section,
the charges were strictly linked to the polymer backbone
and no motion of the charges was allowed.
The only way of minimizing the free energy
was to change the conformation.
With annealed polyelectrolytes, we add an extra degree of freedom and
the trumpet effect is less pronounced.
We also expect
the charges to accumulate preferably at the ends of the chain
since the charges miss there approximately
half of their neighbors compared to the chain center.

In order to be more quantitative,
we model the charges of the chain
as a one dimensional gas of interacting particles along the polymer,
in contact with a charge reservoir which imposes on average the local charge on a chain.
The chemical potential of the charges is directly related to the pH of the solution
\cite{raphael_joanny}.\\
Following the results of the appendix, we write the free energy of the system:

\begin{eqnarray}
\label{f_energy1}
F\left[z(s)\,,\,f(s)\right] & = &
\int_{-N/2}^{N/2}ds\,
\Bigg\{
\frac{3}{2b^{2}}\left(\frac{dz}{ds}\right)^{2}
\nonumber\\
& & +\frac{l_{B}}{2}
\int_{-N/2}^{N/2}ds'\,f(s)\,f(s')
\left\langle
\frac{1}{\left|\mathbf{r}(s)-\mathbf{r}(s')\right|}
\right\rangle_{t}
\nonumber\\
& & + f(s)\,(\ln{f(s)}-1)\,-\,\mu_{\mathrm{ch}}\,f(s)
\Bigg\}
\end{eqnarray}
where $f(s)$ is the local charge distribution along the chain, and $\langle...\rangle_t$ 
refers to the average over the conformation of a chain under tension. The first two terms of Eq.(\ref{f_energy1})
represent the free energy of a quenched polyelectrolyte
with an inhomogeneous charge distribution $f(s)$.
The third term is the entropy of an unidimensional ideal gas
and the last term fixes the charge of the chain.
The minimization of the free energy with respect to $f(s)$
gives a Boltzmann law for  the charge distribution:
\begin{equation}
\label{boltzmann}
f(s)=f(0)\,\exp{\left[-(\phi(s)-\phi(0))\right]}
\end{equation}
with $\phi(s)=\frac{e\varphi(s)}{kT}$,
$\varphi(s)$ being the average electrostatic potential.
On the other hand,
the minimization with respect to  $z(s)$
leads to an equation  for the blob size.
We rewrite those equations in the following way:

\begin{eqnarray}
\label{potential}
\widetilde{\phi}(z) & = &
\frac{l_{B}f(0)}{b^{2}}
\int_{-L/2}^{L/2}dz'\,\xi_{z'}\,\exp{\left(-\widetilde{\phi}(z')\right)}\,
\left(\frac{1}{\left| z-z'\right|}
\quad -\quad\frac{1}{\left|z'\right|}\right)\\
\label{blob}
\xi_{z} & = &
\frac{\xi_{0}}
{\left[1\,+\,\frac{\xi_{0}^{2}}{B^{2}}
\left(1-\exp{\left(-\widetilde{\phi}(z)\right)}\right)
\right]^{1/2}}
\end{eqnarray}
with $B=\frac{3b^{2}}{4f(0)}$ and $\widetilde{\phi}(z)=\phi(z)-\phi(0)$.
 The blob size $\xi_{0}$ is
evaluated at the middle of the chain.
The integrals have to be regularised by short distances cutoffs.\\
Those two equations describe the coupling
between the charge distribution
and the conformation.
The reduced potential $\widetilde{\phi}(z)$ vanishes for an infinite chain
length
and must be small in a wide range of $z$ values
for a finite chain length; this is not true when one approaches
the end of the chain.
As a first approximation,
we solve Eq.(\ref{potential})
by iterating it once with $\widetilde{\phi}(z)$.
We get a first order solution with respect to the parameter
$\frac{l_{B}f(0)\xi_{0}}{b^{2}}$:
\begin{equation}
\label{sol_potential}
\widetilde{\phi}(z)=\frac{l_{B}f(0)\xi_{0}}{b^{2}}\,
\ln{\left(1-\left(\frac{z}{L/2}\right)^{2}\right)}
\end{equation}
This approximation amounts to calculating
the electrostatic potential of a chain
of homogeneous blobs
having all the same size.
The parameter $f(0)$ is the monomer charge fraction
at the middle of the chain,
we replace it by the average charge fraction
\cite{berghold_seidel}:
\begin{equation}
\label{f_average}
\langle f\rangle=\frac{1}{L}\int_{-L/2}^{L/2}dz\,f(z)
\end{equation}
We obtain finally for the charge distribution:
\begin{equation}
\label{f(z)_salt_free}
\frac{f(z)}{\langle f\rangle}=1-\frac{l_{B}\langle
f\rangle\xi_{\mathrm{scaling}}}{b^{2}}
\,\left\{
\ln{\left(1-\left(\frac{z}{L/2}\right)^{2}\right)}
\,+\,
2(1-\ln{2})
\right\}
\end{equation}
Notice that we have replaced $\xi_{0}$ by $\xi_{\mathrm{scaling}}$.\\
Since the charge distribution $f(z)$ depends only on the ratio
$\frac{z}{L}$ in this first order theory for the chain of blobs,
one can express it easily as a function of the contour length $s$
by replacing $\frac{z}{L}$ by $\frac{s}{N}$. The formula Eq.(\ref{f(z)_salt_free}) 
is sketched in figure 2 for different charge fractions.\\
The expansion parameter $\frac{l_{B}\langle
f\rangle\xi_{\mathrm{scaling}}}{b^{2}}$  is the Manning parameter
for the condensation of counterions
on a chain of homogeneous electrostatic blobs (see rescaling below).
In fact, our result is the same as the one that would obtain
at fixed conformation, i.e.
with a chain of homogeneous blobs.
This is true at the first order of the theory
because the end effect on the conformation is weak.
This suggests the calculation of the charge distribution of
a fully stretched chain,
 with a rodlike conformation, with the same average charge fraction.
One finds:
\begin{equation}
\label{f(z)_salt_free_2}
\frac{f(z)}{\langle f\rangle}=1-\frac{l_{B}\langle f\rangle}{b}
\,\left\{
\ln{\left(1-\left(\frac{z}{L/2}\right)^{2}\right)}
\,+\,
2(1-\ln{2})
\right\}
\end{equation}
We obtain therefore the important result
that we can get the charge distribution of a chain of blobs
simply by applying the following scaling or coarse-graining relations
on the charge distribution of a fully stretched chain:

\begin{eqnarray}
\label{scaling1}
b & \longrightarrow & \xi_{\mathrm{scaling}}\\
\langle f\rangle & \longrightarrow & \langle f\rangle \,g
=\langle f \rangle\left(\frac{\xi_{\mathrm{scaling}}}{b}\right)^{2}\\
z=sb & \longrightarrow & \frac{s}{g}\,\xi_{\mathrm{scaling}}
=sb\left(\frac{\xi_{\mathrm{scaling}}}{b}\right)^{-1}\\
L=Nb & \longrightarrow & \frac{N}{g}\,\xi_{\mathrm{scaling}}
=Nb\left(\frac{\xi_{\mathrm{scaling}}}{b}\right)^{-1}
\end{eqnarray}
Those relations are valid at the first order of the theory.
Up to this order,
one can generalize this result to the case of polyelectrolyte in a good solvent
in a straightforward way with the following relations,

\begin{eqnarray}
\label{scaling2}
b & \longrightarrow & \xi_{\mathrm{scaling}}\\
\langle f\rangle & \longrightarrow & \langle f\rangle \,g
=\langle f \rangle\left(\frac{\xi_{\mathrm{scaling}}}{b}\right)^{1/\nu}\\
L=Nb & \longrightarrow & \frac{N}{g}\,\xi_{\mathrm{scaling}}
=Nb\left(\frac{\xi_{\mathrm{scaling}}}{b}\right)^{1-1/\nu}
\end{eqnarray}
with
$\nu =1/2$  for a gaussian chain and
$\nu =3/5$ for a chain with excluded volume.
In this last case, $\xi_{\mathrm{scaling}}\sim \frac{b}{(uf^{2})^{3/7}}$.
The validity of this first order theory
is controlled by the parameter
$\omega=\frac{l_{B}\langle f\rangle\xi_{\mathrm{scaling}}}{b^{2}}$
for the theta solvent case.
If we use for example some typical parameters
of weak polyelectrolytes,
$\langle f\rangle\sim 1/20$, $l_{B}\sim b\sim 7$ \AA\,
, then $\omega=0.4$ and our expression should be a quite
good approximation
for the charge distribution.\\
A better than perturbative treatment to solve Eqs.(\ref{potential}, \ref{blob})
would be required to obtain a more accurate expression of the charge
distribution.
\section{Charge distribution with screened interactions}
\setcounter{equation}{0}\renewcommand{\theequation}{\thesection.\arabic{equation}}
\label{salt}
We investigate in this section the effect
of added monovalent salt in the solution
on the charge distribution along the chain.
The salt does not take part in the
dissociation-recombination of the charges and the counterions:
we assume therefore that
the only effect of salt is to
screen the electrostatic interactions.
This may be a very rough approximation
but it is the simplest way to take
into account the salt
and it allows us to get an insight in
the charge distribution.
The electrostatic potential created by an elementary charge reduces to:
\begin{equation}
\label{screened_potential}
\varphi(r)=\frac{l_{B}kT}{e}\,\frac{\exp{(-\kappa r)}}{r}
\end{equation}
The Debye-H\"uckel screening length is given by
$\kappa ^{2}=8\pi l_{B}c_{s}$,
$c_{s}$ being the concentration of monovalent salt.
For an infinite quenched polyelectrolyte,
the classical picture
is a wormlike chain with a persistence length $l_{P}$
\cite{barrat_joanny_persistence}.
There are two contributions to this length:
the bare persistence length which comes from the intrinsic stiffness
of the neutral backbone,
and the electrostatic persistence length due to the screened interactions
between the charges that stiffen the chain.
The relation between
the screening length and this electrostatic persistence length
is a controversial problem which has been discussed extensively for a couple of decades
since the early work of Odijk \cite{odijk}.
In this paper,
we focus on  a fully flexible chain
where we have $\xi_{\mathrm{scaling}}\gg b$.
For this regime, the most recent results \cite{netz,ha_thirumalai}
seem to confirm the result of
Khokhlov and Khachaturian
\cite{K_K}
\,$l_{P_{e}}\sim \kappa ^{-2}$. We do not need here the precise value of
the persistence length we
will only assume that
\begin{equation}
\label{ineq_blob_kappa}
\xi_{\mathrm{scaling}}\ll \kappa^{-1}\ll l_{P}
\end{equation}

If $\xi_{\mathrm{scaling}}\gg \kappa^{-1}$,
 the chain has a swollen conformation
with an excluded volume parameter of order $\kappa ^{-2}$,
but in this case the inhomogeneity in the charge distribution is small
as the electrostatic interactions are screened
at length scales where they are not yet relevant.
\\
In order to derive the charge distribution, we describe the polyelectrolytes
with a modified version of the "chain under tension model" proposed by
Barrat and Joanny
\cite{barrat_joanny_persistence}.
The original model assumes that
the chain is locally stretched with a tension
whose orientational correlations extend over distances of order $l_{P}$:
at short length scales,
the polymer behaves like a chain of stretching blobs
while at large length scales
it shows the conformation of a wormlike chain.
Here we focus on short distances
since the charge distribution is mainly
influenced by electrostatic interactions at these length scales.
We neglect therefore the interaction between charges far apart along the chain.
The main effect of those interactions is to induce
 excluded volume statistics at large length scales.

One proceeds as before and
writes the free energy for the theta solvent case:

\begin{eqnarray}
\label{energy_salt}
\frac{F[\mathbf{r}_{0}(s)]}{kT}
& = &
\int_{-N/2}^{N/2}ds
\Bigg\{\frac{3}{2b^{2}}\left(\frac{d\mathbf{r}_{0}(s)}{ds}\right)^{2}
\nonumber\\
& & \,+\,
\frac{l_{B}}{2}\int_{-N/2}^{N/2}ds'\,f(s)\,f(s')\,
\left\langle
\frac{\exp{\left(-\kappa \left|\mathbf{r}(s)-\mathbf{r}(s')\right|\right)}}
{\left|\mathbf{r}(s)-\mathbf{r}(s')\right|}
\right\rangle_{t}
\nonumber\\
& & \,+\,
f(s)\left(\ln{f(s)}\,-\,1\right)
\,-\,\mu_{\mathrm{ch}}f(s)
\Bigg\}
\end{eqnarray}
Here $\mathbf{r}_{0}$ denotes the classical path of the polymer
which follows from our simplified model of chain under tension.
After minimization with respect to $f(s)$,
we obtain a Boltzmann law for the charge distribution
that we rewrite it in terms of electrostatic potential only:
\begin{equation}
\label{potential_salt}
\phi (s)=\exp{\left(\frac{\mu_{\mathrm{ch}}}{kT}\right)}
\,l_{B}\,
\int_{-N/2}^{N/2}ds\,\exp{\left(-\phi (s')\right)}
\,\left\langle
\frac{\exp{\left(-\kappa\left|\mathbf{r}(s)-\mathbf{r}(s')\right|\right)}}
{\left|\mathbf{r}(s)-\mathbf{r}(s')\right|}
\right\rangle_{t}
\end{equation}
The use of this equation by itself without using the minimization equation with
respect to the chain conformation
allows the determination of the charge distribution
at a fixed conformation of the polymer. We will use here the average
conformation
of an infinite chain.
This approximation is justified by the considerations of section
\ref{conformation} about the end effect on the conformation
of a polyelectrolyte with a frozen charge distribution.
Here, the number of charges implied in the stretching
of the chain, hence the stretching force, are reduced because of the screened interactions.
The effect on the conformation should be weaker
than in the case of a salt-free solution.\\
We evaluate the average over the chain conformation in Eq.(\ref{potential_salt})
by assuming that the chain is stretched
over distances smaller than $l_{P}$
by a uniform tension $\zeta kT\,\mathbf{t}$,
$\mathbf{t}$ being a unit vector.
This average is first expressed in terms of its Fourier transform:
\begin{equation}
\label{fourier}
\left\langle
\frac{\exp{\left(-\kappa\left|\mathbf{r}(s_{1})-\mathbf{r}(s_{2})\right|
\right)}}
{\left| \mathbf{r}(s_{1})-\mathbf{r}(s_{2}) \right|}
\right \rangle_{t}
=
\int\frac{d^{3}q}{(2\pi)^{3}}
\,\frac{4\pi}{q^{2}+\kappa ^{2}}
\,\left\langle
\exp{\left[i\,\mathbf{q}.\left(\mathbf{r}(s_{1})-\mathbf{r}(s_{2})\right)\right]}
\right\rangle_{t}
\end{equation}
For distances smaller than $l_{P}$, we have:
\begin{equation}
\left\langle
\exp{\left[i\,\mathbf{q}.\left(\mathbf{r}(s_{1})-\mathbf{r}(s_{2})\right)\right]}
\right\rangle_{t}
=\frac{\sin{\left(\frac{q\zeta b^{2}}{3}(s_{1}-s_{2})\right)}}
{\frac{q\zeta b^{2}}{3}(s_{1}-s_{2})}
\,\exp{\left[-\frac{q^{2}b^{2}}{6}(s_{1}-s_{2})\right]}
\end{equation}
after averaging over the direction of stretching.\\
Finally, we get
\cite{gradstein}

\begin{eqnarray}
\label{complex_eq}
\left\langle
\frac{\exp{\left(-\kappa\left|\mathbf{r}(s_{1})-\mathbf{r}(s_{2})\right|\right)}}
{\left|\mathbf{r}(s_{1})-\mathbf{r}(s_{2})\right|}
\right\rangle_{t}=\,-\frac{1}{2}\frac{\exp{(\kappa ^{2}r_{s_{1}s_{2}}^{2})}}{z_{s_{1}s_{2}}}\times\hskip6truecm&\cr
\Bigg\{
\exp{(\kappa z_{s_{1}s_{2}}})\,\left(1-\mathrm{Erf}\left[\kappa
r_{s_{1}s_{2}}+\frac{1}{2}\frac{z_{s_{1}s_{2}}}{r_{s_{1}s_{2}}}\right]
\right)\hskip1truecm&\cr
\quad -\exp{(-\kappa z_{s_{1}s_{2}}})\,
\left(
1-\mathrm{Erf\left[\kappa
r_{s_{1}s_{2}}-\frac{1}{2}\frac{z_{s_{1}s_{2}}}{r_{s_{1}s_{2}}}\right]}
\right)
\Bigg\}&
\end{eqnarray}

We introduced in Eq.(\ref{complex_eq})
the following notations:
$\xi_{0}=1/\zeta$ is the stretching blob size.
$z_{s_{1}s_{2}}=\frac{b^{2}}{3\xi_{0}}(s_{1}-s_{2})$ is the stretched
radius associated
with the contour length $(s_{1}-s_{2})$ and
$r_{s_{1}s_{2}}=\left(\frac{b^{2}}{6}(s_{1}-s_{2})\right)^{1/2}$
the gaussian radius of the same chain segment.
The size of a given chain segment is the gaussian radius for distances
inside a blob while at larger length scales,
it is given by the stretched radius.
For $\xi_{0}\ll z_{s_{1}s_{2}}\ll l_{P}$,
we recover the scaling result for the chain of blobs:
\begin{equation}
\label{effective_potential}
\left\langle
\frac{\exp{\left(-\kappa\left|\mathbf{r}(s_{1})-\mathbf{r}(s_{2})\right|\right)}}
{\left|\mathbf{r}(s_{1})-\mathbf{r}(s_{2})\right|}
\right\rangle_{t}
=
\frac{\exp{\left(-\frac{\kappa b^{2}}{3\xi_{0}}\left|s_{1}-s_{2}\right|\right)}}
{\frac{ b^{2}}{3\xi_{0}}\left|s_{1}-s_{2}\right|}
\end{equation}
One can see in Eq.(\ref{effective_potential})
that the screening length is still the characteristic decay length of the interaction,
so that two charges separated by one persistence length
($l_{P}\gg\kappa^{-1}$) interact only weakly.
\\
According to this result and to our preceding assumption
of neglecting interactions between charges far apart along the chain,
we use Eq.(\ref{effective_potential})
as an interpolating formula over the whole range of length scales,
the short distance divergence being removed by a cutoff at a size
$\xi_{0}$.
In other words,
since charges do not feel each other at distances larger
than the persistence length,
the charge distribution should be the same within our approximations
as the one of a chain of blobs with
an imposed rodlike conformation,
where the charges interact via a screened Coulomb potential.
The conformation of the chain at large lengthscales
is not relevant for the charge distribution.
Therefore, one can compute easily a first order solution
in $\langle f\rangle l_{B}\xi_{0}/b^{2}$
for the electrostatic potential and for the charge distribution:
\begin{equation}
\label{f_salt_blob}
\frac{f(s)}{\langle f\rangle}
=
1\,+\,
\frac{3\langle f\rangle l_{B}\xi_{0}}{b^{2}}
\left\{
\mathrm{E_{1}}\left(\left(\frac{N}{2}+s\right)\widetilde{\kappa}\right)+
\mathrm{E_{1}}\left(\left(\frac{N}{2}-s\right)\widetilde{\kappa}\right)
-\frac{2}{N\widetilde{\kappa}}
\right\}
\end{equation}
with the exponential integral
$\mathrm{E_{1}}(x)=\int_{x}^{+\infty}dt\,\frac{\exp{(-t)}}{t}$
and the screening length expressed in terms of contour length
$\widetilde{\kappa}=\frac{\kappa b^{2}}{3\xi_{0}}$.
The charge distribution is sketched in figure 3 for different screening length.
We have introduce in Eq.(\ref{f_salt_blob})
the average charge fraction $\langle f\rangle$
instead of the parameter $\mu_{\mathrm{ch}}$.
The characteristic decay length of the charge variation
from the ends is of order $\kappa^{-1}$.
\\
As in the preceding section,
it is interesting to compare this result
to the case of a fully stretched chain with the same
average charge :
\begin{equation}
\label{f_salt}
\frac{f(s)}{\langle f\rangle}
=
1\,+\,
\frac{\langle f\rangle l_{B}}{b}
\left\{
\mathrm{E_{1}}\left(\left(\frac{N}{2}+s\right)b\kappa \right)+
\mathrm{E_{1}}\left(\left(\frac{N}{2}-s\right)b\kappa \right)
-\frac{2}{Nb\kappa}
\right\}
\end{equation}
This form for the charge distribution of a fully stretched chain
was first derived by Berghold et al.
\cite{berghold_seidel}.
We obtain therefore the same physical result as in the case of a salt-free
solution:
the charge distributions
Eqs.(\ref{f_salt_blob}, \ref{f_salt})
are obtained from one another by the following
rescaling:

\begin{eqnarray}
b & \longrightarrow & \xi_{0}\\
\langle f\rangle & \longrightarrow & \langle f\rangle\,g\,=\,\langle f\rangle\,
\left(\frac{\xi_{0}}{b}\right)^{1/\nu}\\
sb\,\kappa & \longrightarrow & \frac{s}{g}\,\xi_{0}\,\kappa\,=\,sb\,\kappa
\left(
\frac{\xi_{0}}{b}\right)^{1-1/\nu}\\
L=Nb & \longrightarrow & \frac{N}{g}\,\xi_{0}\,=\,Nb\,\left(
\frac{\xi_{0}}{b}\right)^{1-1/\nu}
\end{eqnarray}
where $\xi_{0}$ is the scaling electrostatic blob size.
As discussed in the preceding section,
those relations are valid both for a theta solvent ($\nu =1/2$)
and a good solvent ($\nu=3/5$) up to the first order of the theory.
We recover the same other qualitative conclusions as in the case of
a salt-free solution.

Note that our results have been derived for
polyelectrolytes that fulfill the inequality
Eq.(\ref{ineq_blob_kappa}).
Otherwise, one has to take more carefully
into account the conformation of the chain.
In particular,
if $\xi_{0}\sim\kappa^{-1}$,
one reaches the so-called cross-over regime
for the persistence length
and a  more sophisticated model is needed.
We believe however that
the characteristic length of the variation of the charge distribution
will still be the Debye-H\"uckel screening length.

One may be tempted to interpret the simulation results
of Berghold et al. \cite{berghold_seidel}
in the framework of our theory.
However, as mentioned in the introduction,
they have not enough charges in their system
to reach the asymptotic regime that we are describing.
For example, the structure factor of the chains
does not evolve when the charge fraction varies
and is characteristic of a chain with excluded volume
interactions.
This indicates clearly that the charges do not perturb the conformation
of this weakly charged polymer in a good solvent
but that they interact anyway with each other in this
highly fluctuating regime corresponding to a
neutral polymer behavior.
The charge inhomogeneity predicted by our model is stronger than that of
the simulation. This is because their situation corresponds to a single electrostatic blob.
The effect should be enhanced if one reaches the many blob regime
described in this paper.
\section{Charge distribution in a poor solvent}
\setcounter{equation}{0}
\label{poor}
The distinction between quenched and annealed polyelectrolytes
is particularly important in a poor solvent where it leads to an unusual
behavior
of titration curves of polyacids or polybases.
This was first studied theoretically by Raphael and Joanny in their
early work on annealed polyelectrolytes \cite{raphael_joanny}.
The exchanges between the charges on the chain
and those in the solution combined with the attractive interactions in the
poor solvent
lead to a charge instability
and to a first order phase transition
between spherical globules and elongated chains of electrostatic blobs. Throughout this section, the
``electrostatic blob'' length designs the lengthscale below which the gaussian statistics 
is only slightly perturbed by the charges. Our aim is to find the charge distribution
along the chains when they are in the stable regime
and to investigate more precisely the relationship
between the charge distribution and the transition
towards the unstable regime.
In their work on annealed polyelectrolytes,
Raphael and Joanny chose a globular blob model
to describe the polyelectrolyte chain in a poor solvent:
at short length scales the attractive interactions
dominate over the electrostatic interaction
while at larger length scales the electrostatic interactions dominate.
Note that at very short length scales,
the attractive interactions are not yet relevant
and are dominated by the gaussian fluctuations of the polymer backbone.
This introduces a thermal blob size $\xi_{T}=b/\tau$,
where $\tau$ is the reduced temperature $\frac{|T-\theta|}{\theta}$
that measures the solvent strength.
The polyelectrolyte conformation can then be described
as a chain of globular blobs,
the internal structure of a globular blob
being a close packing of thermal blobs of size $\xi_{T}$.
The monomer density in this collapsed state is given by
$\tau /b^{3}$. However, this state is always unstable with respect
to the total length fluctuations of the chain because of the weakness of 
the restoring force to stretching of a cylindrical globule. Dobrynin et al. \cite{dobrynin_rubinstein}
have shown that a necklace conformation is more stable, where the chain can be viewed
as a succession of beads
connected by narrow strings (Fig.4).
The beads have the same structure as the globular blobs,
while the strings are described as linear arrays of thermal blobs.
We now  show that the qualitative description of the instability of
annealed polyelectrolytes
with respect to charge fluctuations is still correct
if we start from a necklace conformation
instead of a cylindrical globule conformation.
We neglect therefore in a first step all the charge inhomogeneities
along the chain.\\
The free energy of a quenched polyelectrolyte
in a cylindrical globule conformation of length $L$ and width $D$
has two contributions:
the surface energy $F_{\mathrm{surf}}^{\mathrm{(cyl)}}=\gamma LD$,
$\gamma$ being the surface tension of the globule given by
$\frac{kT}{\xi_{T}^{2}}$,
and the electrostatic energy
$F_{\mathrm{elec}}^{\mathrm{(cyl)}}=\frac{kTl_{B}(fN)^{2}}{L}$
up to a logarithmic factor.
At equilibrium, those energies are of the same order of magnitude
and this gives the value of $D\sim\frac{b}{(uf^{2})^{1/3}}$
with as before $u\equiv l_{B}/b$.
The cylindrical globule conformation
does not exist at low and high values of the charge fraction $f$;
the stable conformations are respectively
the spherical globules and the elongated chain of electrostatic blobs.
By comparing the free energies of those different conformations,
we get the range of stability of the cylindrical globule:
\begin{equation}
\label{globule}
\left(\frac{\tau}{uN}\right)^{1/2}<f<\tau^{3/2}\,u^{-1/2}
\end{equation}
For a quenched polyelectrolyte in a necklace conformation,
we recover two analogous contributions
to the free energy $F_{\mathrm{surf}}^{\mathrm{(nec)}}$ and
$F_{\mathrm{elec}}^{\mathrm{(nec)}}$.
At the level of the scaling laws,
one has $F_{\mathrm{surf}}^{\mathrm{(cyl)}}\sim
F_{\mathrm{surf}}^{\mathrm{(nec)}}$
since the surface of the strings is negligible compared
to that of the beads.
The gain in free energy when one passes
from the cylindrical globule to the necklace comes from the change in
electrostatic interactions:
in the necklace conformation,
the beads are further apart.
Therefore the free energy is slightly reduced
for a necklace.
Following the work of Dobrynin et al.,
the equilibrium free energy reads:
\begin{equation}
\label{energy_poor1}
\frac{F_{\mathrm{quenched}}^{\mathrm{(nec)}}}{kT}=(N-M_{\mathrm{st}})\tau
u^{1/3}\,f^{2/3}
\,+\,M_{\mathrm{st}}\frac{u}{\tau}f^{2}\,+\,
Nf(\tau u)^{1/2}
\end{equation}
In Eq.(\ref{energy_poor1}),
$M_{\mathrm{st}}=fN\left(\frac{u}{\tau ^{3}}\right)^{1/2}$
is the total number of monomers belonging to strings.
The first term is the self-energy of the beads.
The second term is the self-energy of the strings.
Finally, the last term represents the dominant electrostatic interactions, namely the interaction between beads. It is convenient to express each term of
Eq.(\ref{energy_poor1})
as a function of the ratio $\frac{\xi_{T}}{D}$,
$D$ being the diameter of the beads equal to the globular blob size
in the model of Dobrynin et al.
This ratio is small for a typical necklace conformation and allows us
to compare the relative magnitude of the various terms
in Eq.(\ref{energy_poor1}):

\begin{eqnarray}
\label{magnitude}
N\tau u^{1/3}\,f^{2/3}
& \sim & N\tau^{2}\frac{\xi_{T}}{D}\\
Nf(\tau u)^{1/2}
& \sim & N\tau^{2}\left(\frac{\xi_{T}}{D}\right)^{3/2}\\
M_{\mathrm{st}}\tau u^{1/3}\,f^{2/3}
& \sim & N\tau^{2}\left(\frac{\xi_{T}}{D}\right)^{5/2}\\
M_{\mathrm{st}}\frac{uf^{2}}{\tau}
& \sim & N\tau^{2}\left(\frac{\xi_{T}}{D}\right)^{9/2}
\end{eqnarray}
Some of the terms of the free energy
are not relevant, but we keep them since they help in the interpretation of some of the following results.
\\
The free energy of an annealed polyelectrolyte in a necklace conformation reads:
\begin{equation}
\label{energy_poor2}
\frac{F_{\mathrm{annealed}}^{\mathrm{(nec)}}}{kT}
=\frac{F_{\mathrm{quenched}}^{\mathrm{(nec)}}}{kT}
+Nf\,(\ln{f}-1)\,-\mu_{\mathrm{ch}} N f
\end{equation}
where we add the entropy of the charges along the chain
and the chemical potential associated to exchanges between the charges and
the solution.
The equilibrium value of the average charge fraction $f$ minimizes
the free energy Eq.(\ref{energy_poor2}):
\begin{equation}
\label{chemical_pot}
\frac{\mu_{\mathrm{ch}}}{kT}
=\frac{\mu (f)}{kT}
=
\ln{f}\,+\,\frac{2}{3}\frac{(N-M_{\mathrm{st}})}{N}\frac{\tau u^{1/3}}{f^{1/3}}
\,+\,2\frac{M_{\mathrm{st}}}{N}\frac{u}{\tau}f
\,+\,(\tau u)^{1/2}
\end{equation}
The charge chemical potential,
is equal up to a constant to the pH of the solution; it
fixes the average charge of the chain.
In titration experiments,
one plots $\mu (f)$ vs $f$.
The charge fraction obtained from Eq.(\ref{chemical_pot})
corresponds to a minimum of the free energy if
\begin{equation}
\label{ineq_stable1}
\frac{1}{f}-\frac{2}{9}\frac{(N-M_{\mathrm{st}})}{N}\frac{\tau u^{1/3}}{f^{4/3}}
+2\frac{M_{\mathrm{st}}}{N}\frac{u}{\tau}>0
\end{equation}
Up to lowest order in $\frac{\xi_{T}}{D}$,
this inequality is rewritten as:
\begin{equation}
\label{ineq_stable2}
f^{1/3}>\frac{2}{9}\tau u^{1/3}
\end{equation}
Since the necklace conformation, as the cylindrical globule
 exists for the range of charge fraction
given by Eq.(\ref{globule}), the polyelectrolyte chain in the necklace conformation
is stable with respect to charge fluctuations
if
\begin{equation}
\label{stable}
\tau <u^{-3/5}\,N^{-1/5}
\end{equation}
up to a numerical prefactor.
One recovers the upper bound for stability of the
cylindrical globule phase found by Raphael and Joanny.
This is a trivial result since the dominant term
in the free energies and so the underlying physics
are the same.
Notice the close relation between this instability
and the counterion condensation on a chain with
a quenched charge distribution.
In this last case,
the inequality Eq.(\ref{stable})
indicates the regime where one never has
counterion condensation for a necklace conformation
\cite{dobrynin_rubinstein}
at fixed $\tau $.
Furthermore, Dobrynin et al.
\cite{dobry2}
show in a recent paper that there is no condensation of counterions on the beads
if the inequality Eq.(\ref{ineq_stable2}) holds.
Those analogies come from the fact
that in our formalism for the annealed charge distribution
exchanges between the charges on the chain
and in the solution
are taken into account,
and counterion condensation is also described.

Eq.(\ref{chemical_pot}) and Eq.(\ref{ineq_stable1})
were derived at fixed conformation.
It means that we imposed that $M_{\mathrm{st}}$ is independent of $f$.
If we relax this constraint and take into account the variation of
$M_{\mathrm{st}}$ with
 $f$,
higher order terms are added to
Eq.(\ref{chemical_pot}) and Eq.(\ref{ineq_stable1})
but the final result is not charged.
For $\tau <u^{-3/5}\,N^{-1/5}$,
$\mu (f)$ is an increasing function of $f$,
while for $\tau >u^{-3/5}\,N^{-1/5}$
it shows a decreasing part in a range of $f$ values.
One can show that the stable state for this latter case
is obtained from a Maxwell equal area construction
\footnote{One has to equal the free energy of a spherical globule and of a
chain of blobs to obtain this construction.}
for the curve $\mu (f)$.
This corresponds to two coexisting type of chains
in the solution:
spherical globules with low $f$
and chains of thermal blobs with high $f$.

A way of getting more precise information about the instability
is to investigate the evolution of the charge distribution along the chain
when the system approaches the transition for
$\tau <u^{-3/5}\,N^{-1/5}$
where the necklace conformation is still stable.
We present here a simplified model for the charge distribution:
we assume that each monomer of the beads has a charge
$f-\delta f$ and each monomer of the strings has a charge
$f+\delta f'$,
and we estimate the corresponding free energy.
The choice of the sign of the deviation around the average value $f$
is motivated by the fact that the charges are closer in the beads
than in the strings.
If one allows the charges to move,
we expect therefore a charging of the strings.
The strength of the effect is related to the distance
to the transition as shown below.
Of course this two-phase approximation for the the charge distribution
does not take into account the finite length of the chain as in the
preceding sections
but we expect that the effect of chain ends is small on the transition.
For a sake of simplicity,  we  work as before
at fixed conformation
and impose $M_{\mathrm{st}}$
to be independent of $f$.
A more precise calculation that relaxes this constraint
gives the same result at leading order.
\\
Since we are dealing with a homogeneous scaling theory,
we use the following prescription for the evaluation
of the changes in the free energy Eq.(\ref{energy_poor2}):
\begin{itemize}
\item $f\rightarrow f-\delta f $
in the self-energy of the beads
\item $f\rightarrow f+\delta f' $
in the self-energy of the strings
\item $Nf\rightarrow (N-M_{\mathrm{st}})(f-\delta
f)+M_{\mathrm{st}}(f+\delta f') $
in the interaction energy
between the beads and in the exchange term
\item $Nf\ln{f}\rightarrow (N-M_{\mathrm{st}})(f-\delta f)\ln{(f-\delta f)}
+M_{\mathrm{st}}(f+\delta f')\ln{(f+\delta f')} $\\
in the entropy of the charges
\end{itemize}
In a first approach,
if we assume that the charge fluctuations
$\delta f$ and $\delta f'$
are small,
we can compute the change in the free energy
$\Delta F=F_{\mathrm{annealed}}^{\mathrm{(nec)}}(f,\delta f,\delta f')
-F_{\mathrm{annealed}}^{\mathrm{(nec)}}(f,0,0)$
up to second order in $\delta f$ and $\delta f'$
and we get finally the charge distribution
along the chain by minimizing $\Delta F$
with respect to $\delta f$ and $\delta f'$:

\begin{eqnarray}
\label{delta_f}
\delta f
& = &
\frac{\frac{M_{\mathrm{st}}}{N}
\left(\frac{2}{3}\frac{\tau u^{1/3}}{f^{1/3}}
-2\frac{uf}{\tau}
\right)}
{\frac{1}{f}-\frac{2}{9}\frac{\tau u^{1/3}}{f^{4/3}}}\\
\label{delta_f'}
\delta f'
& = &
\frac{\frac{N-M_{\mathrm{st}}}{N}
\left(
\frac{2}{3}\frac{\tau u^{1/3}}{f^{1/3}}
-2\frac{uf}{\tau}
\right)}
{\frac{1}{f}+2\frac{u}{\tau}}
\end{eqnarray}
where the equilibrium charge fraction $f$ is given by Eq.(\ref{chemical_pot}).

Before interpreting those results,
 we check their consistency with our assumptions.
First the deviations $\frac{\delta f}{f}$ and $\frac{\delta f'}{f}$ are of order
$\left(\frac{\xi_{t}}{D}\right)^{\alpha}$
with $\alpha >1$,
except when the denominator of Eq.(\ref{delta_f})
vanishes,
a case discussed below.
Therefore the effect of the charge distribution between the beads and the strings
is weak.
One can explain the weakness of the effect with the following argument:
inside a bead, the electrostatics perturbs only slightly
the conformation of the chain.
On the opposite, two beads interact strongly
because of their high charges.
If we look for the electrostatic potential produced by
a quenched conformation of charges,
we expect the beads to behave as a potential well:
the first relevant interactions of a test charge in a string
with the neighboring charges are those
with the two closest beads and so occur
at shorter distances than for a test charge in a bead.
In a very rough estimation,
the electrostatic potential in a string is twice that
in a bead.
This explains the weakness of the charging of the strings.
\\
In agreement with what we were expecting,
$\delta f$ and $\delta f'$ are positive.
The first positive term in the numerator of
Eq.(\ref{delta_f}) and Eq.(\ref{delta_f'})
represents the electrostatic energy of the charges
in the beads and tends
to deplete the charges from the beads
while the second term in the numerator represents the electrostatic
energy of the charges in the strings
and tends to limit the charging
of the strings.
The leading term is the first one.

As mentioned before,
a change in the charge distribution of a necklace
drives a change in the conformation.
For example, if one decreases the bead charge,
the bead size increases because  more monomers carrying less charges
can be aggregated.
Since the charge distribution inhomogeneities are weak,
except for the region of instability,
it is a reasonable approximation to work at fixed conformation,
although we were able to manage the calculation including a change in
conformation
where the parameter $M_{\mathrm{st}}$ remains free to adjust at its optimal
value
as mentioned above.

The most striking feature of Eq.(\ref{delta_f})
is that according to the inequality Eq.(\ref{ineq_stable2}),
the deviation of the charge of a monomer in a bead
is diverging when $\tau$ reaches the value
$u^{-3/5}N^{-1/5}$.
Therefore, the charge instability
corresponds to a divergence of the inhomogeneity
in the charge distribution along the chain.
However, our simple calculation
does not allow any quantitative description
of the transition towards coexisting phases
at $\tau \sim u^{-3/5}N^{-1/5}$
since it is  a perturbative treatment.

\section{Concluding remarks}

We have  presented  in this paper
an estimation of the equilibrium charge distribution
on an annealed polyelectrolyte chain under various conditions.
The charge distribution is mostly an effect of the finite length of the chain,
since an infinite chain possess a translational invariance
(for the equilibrium properties).
Our model for a salt-free solution
in a theta or a good solvent
leads to self-consistent equations
coupling the conformation of the chain
and the charge distribution.
The chain conformation is the
so-called classical path of the polymer.
All the fluctuations around
this most probable path are neglected.

When one fixes the distribution of the charges along the chain,
one can investigate the conformation of the chain alone.
We predict in this case a trumpetlike
shape, with a local electrostatic blob size $\xi (z)$ increasing towards
the end of the chains.
The explicit dependency of the blob size
with $z$ is found by using a model very similar to that used by Brochard
to study polymers in strong flows.
The end effect on the conformation is weak and the conformation is very close to the one
obtained from homogeneous scaling arguments.

When the charge distribution is annealed,
we only need to consider an imposed conformation for the chain as
a linear array of electrostatic blobs with all the same size.
The charges accumulate at the ends of the chain.
This result does not change the qualitative behavior
of an annealed polyelectrolyte solution compared to a solution of
polyelectrolytes
with quenched charge distributions,
but it shows that annealed polyelectrolytes might
exhibit some interesting collective properties
where end effects are known to be very important (such as adsorption on a
charged surface).
In the case of theta or good solvent for example,
similar properties have been predicted
\cite{zhulina_borisov}
for  weak polyelectrolyte brushes.
\\
With a similar model,
we show that the effect of added salt
is to produce a faster decay
of the inhomogeneity in the charge distribution along the chain.
This charge heterogeneity decays over a length of the order of the
Debye-H\"uckel screening length.
Our predictions for the charge distribution cannot however be directly compared
to the first simulation made by Berghold et al.
\cite{berghold_seidel}
because we are not describing the same scaling regimes for the charge
distribution.

In the case of a solution in  a poor solvent,
one finds another example of the typical collective properties of the
annealed polyelectrolytes mentioned above: within a certain range
of parameters, there is a first order phase transition
between spherical globules with low $f$
and chain of electrostatic blobs with high $f$,
induced by a charge instability on the chains.
We show that this instability is associated
with the divergence of the charge distribution inhomogeneity along
the chain.

Our work suggests that it would be interesting
to have more precise experimental or numerical evidence
on the behavior of annealed polyelectrolytes in poor solvents.
A comparison with the simulations of Micka et al.
\cite{micka}.
on the behavior of strongly charged
quenched polyelectrolytes in poor solvent
would also give an insight on
the close relation between counterion condensation on quenched polyelectrolytes
and the annealed charge distribution.

{\bf Aknowledgements:} We would like to thank C. Seidel and R. Netz for helpful
discussions about the various behaviors of annealed polyelectrolytes. This
research was supported
by the Deutsche Forschung Gemeinschaft through the Schwerpunkt
"Polyelektrolyte".
%
%
%
%
%
%
%
%
%
\section*{Appendix: Introduction of a blob size with a model of chain
under tension.}
\setcounter{equation}{0}\renewcommand{\theequation}{A.\arabic{equation}}
In this appendix, we show
how the stretching blob size can be interpreted
as a short distance cutoff of the electrostatic interactions
by considering a variational model of chain under tension.
The description of a polyelectrolyte with such a model
has proven successful
(for a review,
see for example
\cite{barrat_joanny_gen})
and it has the important advantage of allowing for explicit calculations.

The hamiltonian of a single chain is:
\begin{equation}
\label{real_hamil}
\frac{H}{kT}
=
\int_{-N/2}^{N/2}ds\,
\left\{
\frac{3}{2b^{2}}
\left(\frac{d\mathbf{r}}{ds}
\right)^{2}
\,+\,
\frac{l_{B}f^{2}}{2}\int_{-N/2}^{N/2}ds'\,
\frac{1}{\left\vert\mathbf{r}(s)-\mathbf{r}(s')\right\vert}
\right\}
\end{equation}
We choose as a trial hamiltonian the one of a chain under a tension
$\mathbf{F}(s)$:
\begin{equation}
\label{trial_hamil}
\frac{H_{t}}{kT}
=
\int_{-N/2}^{N/2}ds\,
\left\{
\frac{3}{2b^{2}}
\left(\frac{d\mathbf{r}}{ds}
\right)^{2}
\,-\,
\mathbf{F}(s).\frac{d\mathbf{r}}{ds}
\right\}
\end{equation}
so that the upper bound of the free energy
reads up to an irrelevant constant:
\begin{equation}
\label{free_energy}
\frac{F_{\mathrm{bound}}(z(s))}{kT}
=
\int_{-N/2}^{N/2}ds\,
\left\{
\frac{3}{2b^{2}}
\left(\frac{dz}{ds}
\right)^{2}
\,+\,
\frac{l_{B}f^{2}}{2}
\int_{-N/2}^{N/2}ds'\,
\left\langle
\frac{1}{\vert\mathbf{r}(s)-\mathbf{r}(s')\vert}
\right\rangle_{t}
\right\}
\end{equation}
where
\begin{equation}
\label{def_z}
\frac{dz}{ds}
\,\equiv\,
\left\langle\left(\frac{d\mathbf{r}}{ds}\right)\right\rangle_{t}
\,.\,
\frac{\mathbf{F}(s)}{\mathrm{F}(s)}
\,=\,
\frac{\mathrm{F}(s)b^{2}}{3}
\end{equation}
The brackets denote an average over the trial statistical weight.
With Eq.(\ref{def_z})
we choose $z(s)$ as the variational parameters instead of $\mathrm{F}(s)$.
\\
To compute the electrostatic contribution to the free energy,
we evaluate the average of
the inverse of the distance between two monomers.
This is done in a straightforward way:
\begin{equation}
\label{average1}
\left\langle
\frac{1}{\left\vert
\mathbf{r}(s)-\mathbf{r}(s')\right\vert}
\right\rangle_{t}
=
\frac{\mathrm{Erf}\left[
\left(\frac{3}{2}\right)^{1/2}
\frac{\left\vert z(s)-z(s')\right\vert}
{\left\vert s-s'\right\vert^{1/2}b}
\right]}
{\left\vert z(s)-z(s')\right\vert}
\end{equation}
with the error function
$\mathrm{Erf}(x)=\frac{2}{\sqrt{\pi}}\,\int_{0}^{x}dy\,\exp{(-y^{2})}$.
For short separations, the average can be rewritten as:
\begin{equation}
\label{average2}
\left\langle
\frac{1}{\left\vert
\mathbf{r}(s)-\mathbf{r}(s')\right\vert}
\right\rangle_{t}\stackrel{s\to s'}{\longrightarrow}
\quad\left(\frac{3}{\pi}\right)^{1/2}\,
\frac{1}{b\left| s-s'\right|^{1/2}}
\end{equation}
There is therefore no short distances divergence
in the calculation of the average electrostatic potential.
This is due to the highly fluctuating state
of a stretched gaussian chain
for length scales smaller than the stretching blob size.
The average of the inverse distances
turn out to be close to the inverse of the average distances
when the argument of the error function
is of order of one.
This defines in a very natural way
the number of monomers in a blob:
\begin{equation}
g=\left(\frac{b}{\alpha}\right)^{2}
\end{equation}
Since a blob has gaussian statistics $\xi\sim g^{1/2}b$,
we recover the definition of the stretching blob size
$\frac{dz}{ds}=\alpha =\frac{b^{2}}{\xi}$.
\\
Therefore, we find that
the charges located in a blob centered at $z(s)$
do not introduce short distances divergence,
and so do not contribute significantly to the average electrostatic
potential at this point.
For reasons of simplicity,
we replace in the rest of the paper
the regularising effect of the error function
implied in the calculation of the average electrostatic potential
by the introduction of short distances cutoffs.
Finally, we notice that we get the same result for the blob size
under the same approximations (weak effect)
as the calculation presented in section
\ref{salt_free}
by minimizing the upper bound of the free energy derived in this appendix.
We actually use this last formalism to obtain the charge distribution along
an annealed polyelectrolyte chain.

%

%
%
\newpage
\section*{Figure Captions}

Figure 1: Trumpetlike conformation:
$L$ is the total length of the chain,
$\xi_{\mathrm{M}}$ is the maximal blob size.
\\
Figure 2: Charge distribution in a salt-free solution under theta solvent
conditions:
$N=130$,
$l_{B}=b=7$\AA .
\\
Figure 3: Charge distribution with added salt under theta solvent conditions:
$\langle f\rangle=1/20$,
$N=130$,
$l_{B}=b=7$\AA. 
\\
Figure 4: Necklace conformation: the inhomogeneity in the charge distribution is modeled by different charging of beads $f-\delta f$ and strings $f+\delta f'$.


\newpage
\begin{figure}[p]
\includegraphics*[width=15cm]{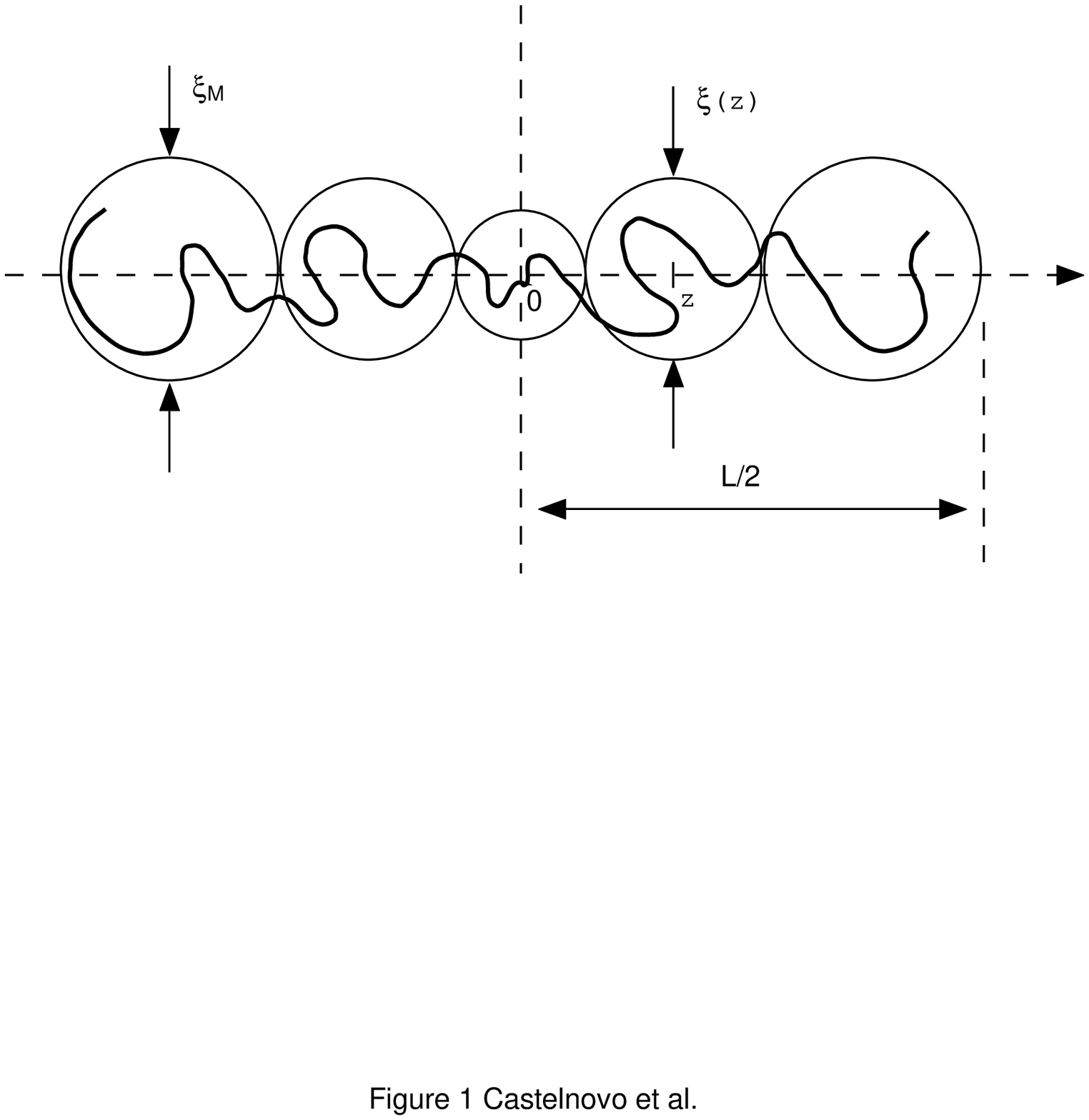}
\end{figure}
\newpage
\begin{figure}[p]
\includegraphics*[width=15cm]{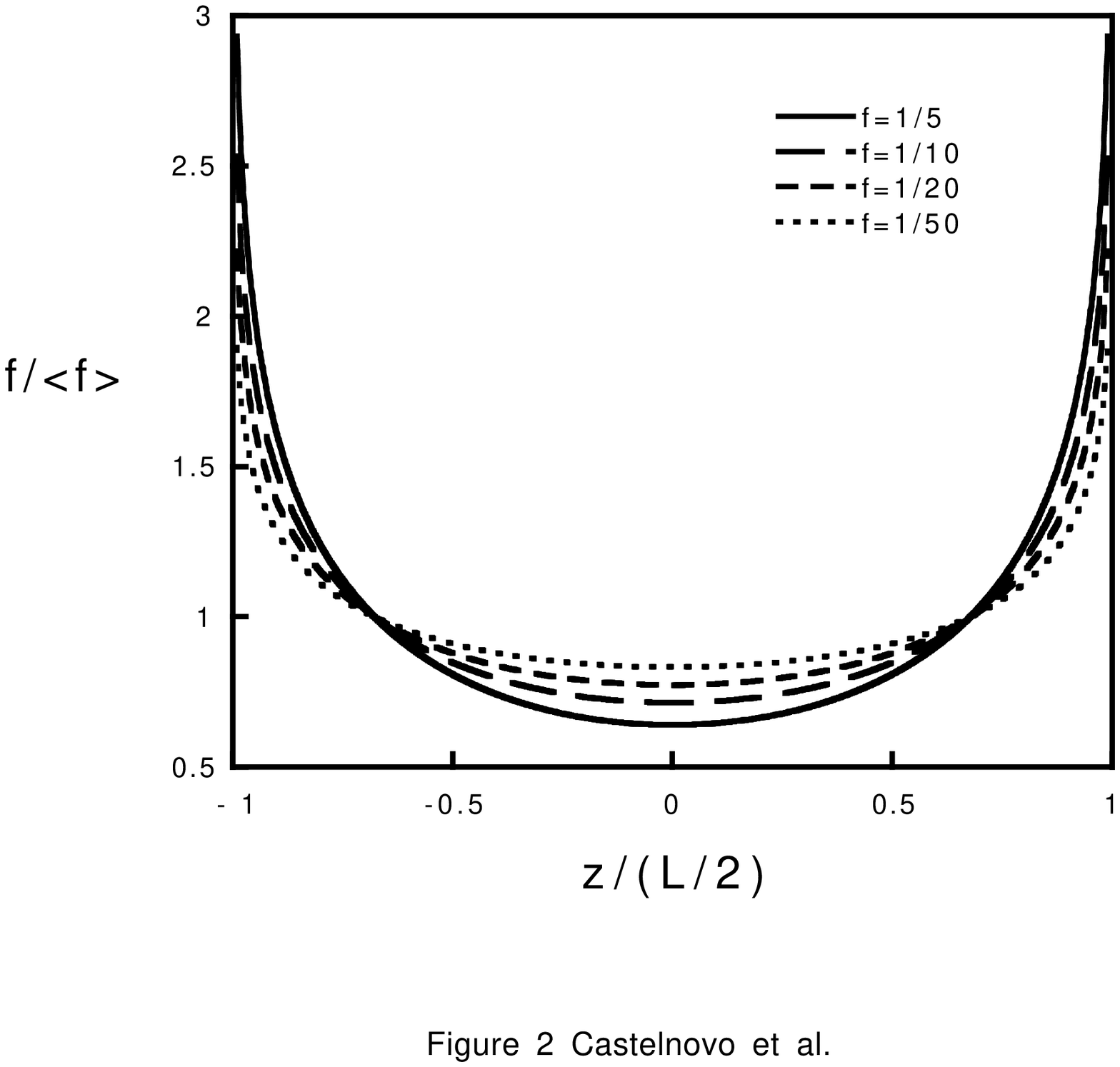}
\end{figure}
\newpage
\begin{figure}[p]
\includegraphics*[width=15cm]{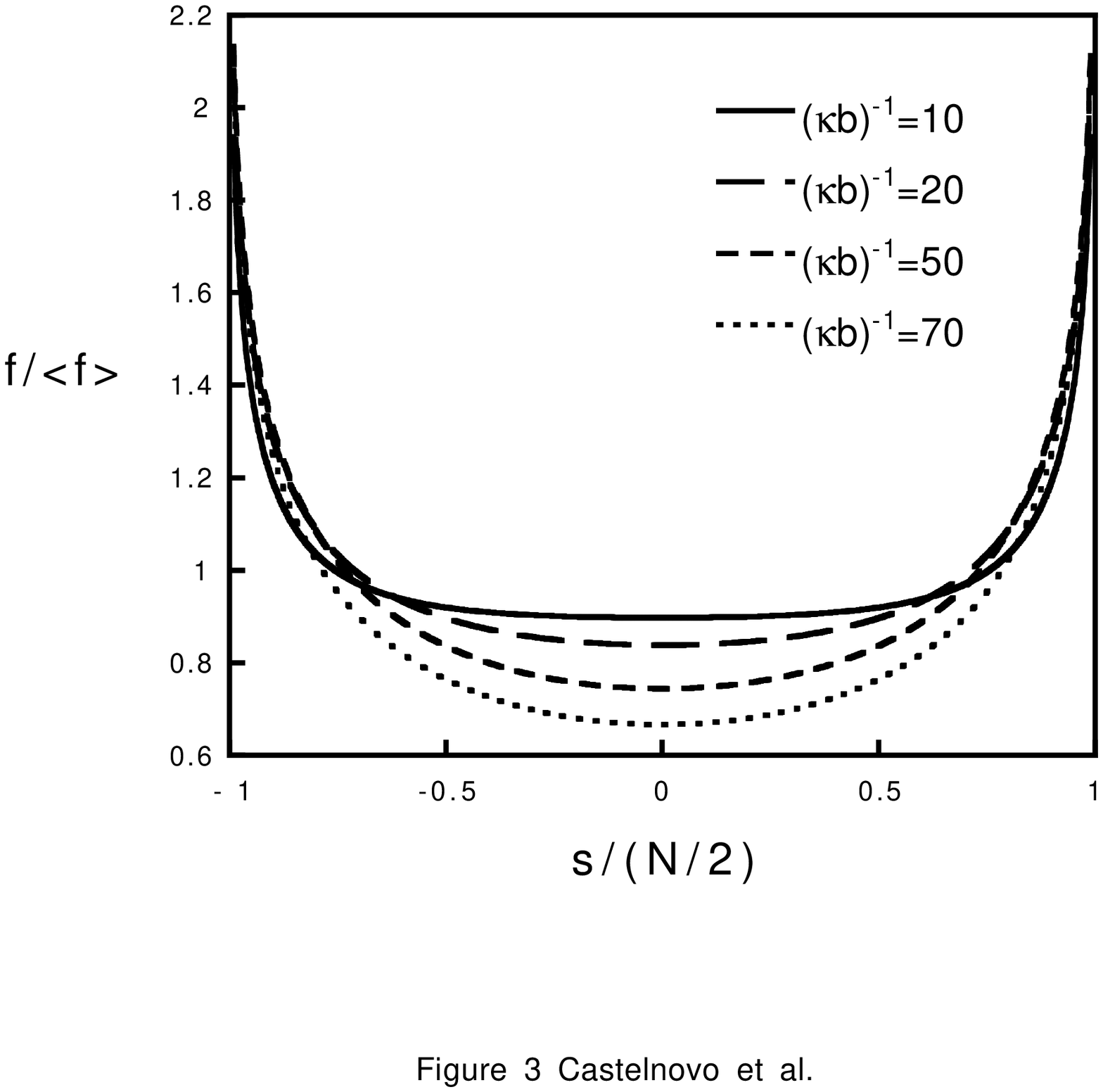}
\end{figure}
\newpage
\begin{figure}[p]
\includegraphics*[width=15cm]{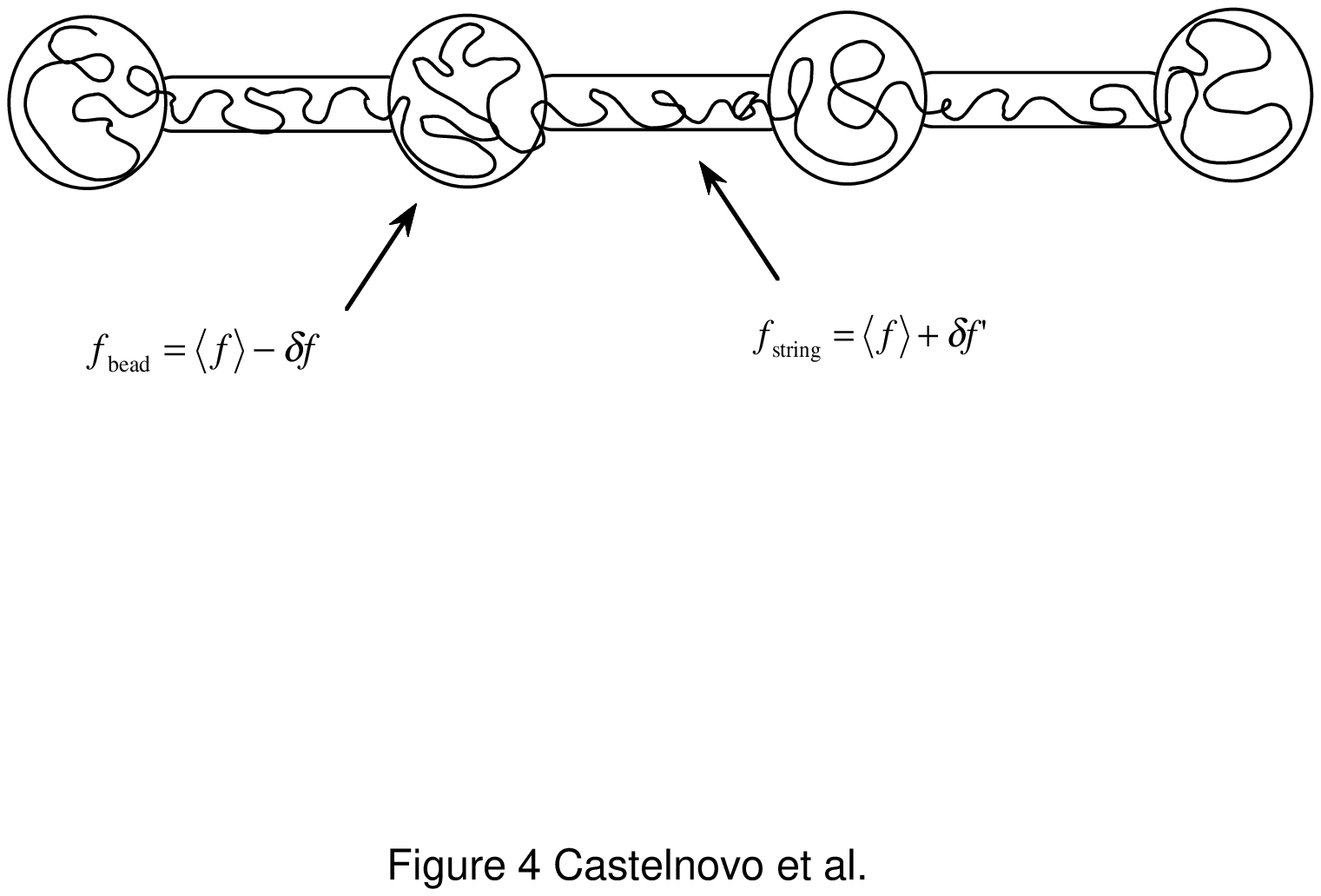}
\end{figure}

\end{document}